\documentclass[a4paper]{spie}  %>>> use this instead for A4 paper
\usepackage{amsmath,amsfonts,amssymb}
\usepackage{graphicx}
\usepackage{astro_bib_macro}

\usepackage[colorlinks=true, allcolors=blue]{hyperref}

\title{Mid-order wavefront control for exoplanet imaging: preliminary characterization of the segmented deformable mirror and Zernike wavefront sensor on HiCAT}

\author[a]{Benjamin Buralli}
\author[a]{Mamadou N'Diaye}
\author[b]{Raphaël Pourcelot}
\author[a]{Marcel Carbillet}
\author[b]{Emiel H. Por}
\author[c]{Iva Laginja}
\author[d]{Ludovic Canas}
\author[b]{Sarah Steiger}
\author[b,e]{Peter Petrone}
\author[b]{Meiji M. Nguyen}
\author[b]{Bryony Nickson}
\author[f]{Susan F. Redmond}
\author[b]{Ananya Sahoo}
\author[b]{Laurent Pueyo}
\author[b]{Marshall D. Perrin}
\author[b]{Rémi Soummer}

\affil[a]{Université Côte d’Azur, Observatoire de la Côte d’Azur, CNRS, Laboratoire Lagrange, France}
\affil[b]{Space Telescope Science Institute, 3700 San Martin Drive, Baltimore, MD 21218, USA}
\affil[c]{LESIA, Observatoire de Paris, Université PSL, Sorbonne Université, Université Paris Cité, CNRS, 5 place Jules Janssen, 92195 Meudon, France}
\affil[d]{Thales Alenia Space, Cannes, France}
\affil[e]{NASA Goddard Space Flight Center, Greenbelt, MD 20771, USA}
\affil[f]{Caltech and Jet Propulsion Laboratory, Pasadena, CA, USA}

\begin{document} 
\maketitle

\begin{abstract}

We study a mid-order wavefront sensor (MOWFS) to address fine cophasing errors in exoplanet imaging with future large segmented aperture space telescopes. Observing Earth analogs around Sun-like stars requires contrasts down to $10^{-10}$ in visible light. One promising solution consists of producing a high-contrast dark zone in the image of an observed star. In a space observatory, this dark region will be altered by several effects, and among them, the small misalignments of the telescope mirror segments due to fine thermo-mechanical drifts. To correct for these errors in real time, we investigate a wavefront control loop based on a MOWFS with a Zernike sensor. Such a MOWFS was installed on the high-contrast imager for complex aperture telescopes (HiCAT) testbed in Baltimore in June 2023. The bench uses a 37-segment Iris-AO deformable mirror to mimic telescope segmentation and some wavefront control strategies to produce a dark zone with such an aperture. In this contribution, we first use the MOWFS to characterize the Iris-AO segment discretization steps. For the central segment, we find a minimal step of 125\,$\pm$31\,pm. This result will help us to assess the contribution of the Iris-AO DM on the contrast in HiCAT. We then determine the detection limits of the MOWFS, estimating wavefront error amplitudes of 119 and 102\,pm for 10\,s and 1\,min exposure time with a SNR of 3. These values inform us about the measurement capabilities of our wavefront sensor on the testbed. These preliminary results will be useful to provide insights on metrology and stability for exo-Earth observations with the Habitable Worlds Observatory.  

\end{abstract}

\keywords{Zernike wavefront sensor, segmented telescope, high-contrast imaging, mid-order wavefront sensor, space adaptive optics.}

\section{Introduction}
\label{sec:intro} 

High-contrast imaging is a promising method to gather information about the physical and chemical properties of planetary companions, i.e. the luminosity, distance to their host star, radius, orbital period, mass, and atmospheric features. On the current facilities, this technique allows us to observe young or massive gaseous exoplanets with a contrast down to $10^{-6}$ down to 200\,mas from their host star in the near infrared. With future observatories, the community aims to image mature or light rocky planets with a contrast down to $10^{-10}$ at angular separations shorter than 50\,mas in the visible. 

To achieve such contrast levels, one exciting solution consists in using a large space telescope with high-contrast capabilities. Missions with large segmented primary mirrors, such as the James Webb Space Telescope (JWST), allow us to obtain the required sensitivity and resolution to observe Earth-like planets. High-contrast capabilities such as the combination of coronagraphy and wavefront control strategies are expected to achieve and stabilize the required contrast. In the past few years, several mission concepts have been studied in the community to enable the observation of terrestrial planets in visible and near-infrared light, such as LUVOIR \cite{Luvoir_2019} and HabEx \cite{Habex_2020}. The 2020 NASA Decadal Survey \cite{Decadal_survey_2021} has proposed the study of Habitable Worlds Observatory (HWO) as a future general astrophysics observatory with high-contrast capabilities to observe a large sample of Earth twins.  

Such an observatory will see its exoplanet imaging capabilities quickly degraded by thermo-mechanical drifts. They will induce different effects and among them, fine misalignments of the telescope primary mirror. These segment cophasing errors will lead to wavefront aberrations and therefore, contrast degradation. To address these telescope segment misalignments in real time, one encouraging solution consists in implementing a dedicated wavefront sensing and control loop.  

In this paper, we consider a wavefront control with a mid-order wavefront sensor (MOWFS) to address the aberrations with mid-order spatial frequency content due to segment misalignments. We first present the MOWFS using a Zernike wavefront sensor and we detail its implementation on HiCAT, the high-contrast testbed for segmented aperture telescopes in Baltimore. This bench is equipped with an Iris-AO segmented deformable mirror to mimic the segmentation of the telescope primary mirror. We perform experimental tests with the MOWFS to characterize the Iris-AO. Finally, the capabilities of the MOWFS are assessed to determine its detection limit in wavefront error amplitude.  

\section{Mid-order wavefront sensor on HiCAT testbed}

\subsection{Zernike wavefront sensor}

For the MOWFS, we consider the use of the Zernike wavefront sensor (ZWFS), since we are interested in controlling small phase aberrations with segmented aperture telescopes. This concept is based on the phase contrast method developed by F. Zernike \cite{Zernike_1934} for microscopy. Its interest in astronomy is much more recent \cite{Angel_1994}. Since then, its use has widely spread among the astronomy community \cite{Bloemhof_2003, Dohlen_2004, Surdej_2010, Vigan_2011, Wallace_2011, Jensen-Clem_2012, N'Diaye_2013, Jackson_2016, N'Diaye_2016, Janin-Potiron_2017, Vigan_2019, Ruane_2020, Steeves_2020, Vigan_2022, Van_Kooten_2022, Salama_2024} for adaptive optics, high-contrast imaging, and picometric metrology.

Similarly to the pyramid wavefront sensor (PWFS) \cite{Ragazzoni_1996}, the ZWFS is a Fourier-filtering wavefront sensor (FFWFS) \cite{Fauvarque_2016} which class of sensors is known for their very high sensitivity. This wavefront sensor uses a focal plane mask with a phase shift of $\theta$ and a diameter of about a resolution element $\lambda/D$, in which $\lambda$ and $D$ denote the wavelength of observation and the diameter of the telescope aperture. Recent studies have showed that the mask diameter can be adjusted to increase the sensor sensitivity \cite{Chambouleyron_2021}.

We briefly recall the principle of this sensor. A point-like source is observed with a telescope aperture located in the entrance pupil plane A. We assume an electric field in this plane with a phase error $\varphi$. In the following focal plane B, the source image is formed and we introduce the Zernike phase mask. The light going through the mask will be phase-shifted and interfere with the light surrounding the mask. This will lead to a pupil intensity $I_C$ in the re-imaged pupil plane C, that is directly related to $\varphi$. From the literature \cite{N'Diaye_2013}, the terms $\varphi$ and $I_C$ are related as follows

\begin{align}
    % I_C = P_A^2 + 2b^2 (1 - \cos \theta) + 2 P_A b (\sin \varphi \sin \theta - \cos \varphi (1 - \cos \theta)) \label{eq1}  \\
    \varphi = \arcsin \left( \frac{I_C - P_A^2 - 2b^2 (1- \cos \theta)}{4 P_A b \sin \frac{\theta}{2}} \right) + \frac{\theta}{2}, \label{eq:phase_general} 
\end{align}

with b the wave diffracted by the mask in the re-imaged pupil plane and $P_A$ the amplitude of the electric field in the entrance pupil plane. In the regime of very small aberrations ($\varphi \ll 1$ rad) and with a typical phase shift $\theta = \pi /2$, the previous equation leads to a linear relation between $\varphi$ and $I_C$ with 

\begin{align}
    \varphi = \frac{I_C}{2 P_A b} - \frac{P_A}{2b} + 1 - \frac{b}{P_A}. \label{eq:phase_linear}
    %\varphi = \frac{1}{\sin \theta} \left( \frac{I_C}{2 P_A b} - \frac{P_A}{2b} + \left( 1 - \frac{b}{P_A}(1 - \cos \theta )\right)\right)
\end{align}

With the ZWFS, we consider two methods for the phase reconstruction, the analytical method and the interaction matrix method. The analytical method uses the equations described above. The interaction method is based on the procedure used in adaptive optics. For a segmented pupil, we introduce a set of piston, tip and tilt for each segment, and we measure the response in the re-imaged pupil plane.  We gather the resulting intensities for all the modes and all the segments into an interaction matrix. Using a singular value decomposition, we pseudo-invert this matrix to obtain the command matrix. This resulting matrix is then used to reconstruct the phase error for a given intensity measured on the Zernike detector. During preliminary simulations, we tested both methods and obtained similar and consistent results. In the following, we will consider the phase reconstruction based on the interaction matrix method for our test with the MOWFS on the HiCAT testbed.

\subsection{High-contrast imager for Complex Aperture Telescopes testbed}

The high-contrast imager for Complex Aperture Telescopes (HiCAT) testbed is an optical bench located at the STScI in Baltimore \cite{Soummer_2024}. It aims to develop high contrast coronagraphic techniques for segmented telescopes, providing an integrated solution for wavefront control and starlight suppression on segmented aperture geometries. The testbed can operate in different coronagraphic modes: Classical Lyot Coronagraph (CLC), Apodized Lyot Coronagraph (APLC) \cite{Soummer_2005}, or Phase-Apodized-Pupil Lyot Coronagraph (PAPLC) \cite{Por_2020}. HiCAT also includes an Iris-AO DM to mimic the pupil segmentation and two continuous kilo-DMs from Boston Micromachines for wavefront control. Different algorithms are available to produce a high-contrast region in the coronagraphic image of the light source. There is also a low-order wavefront control loop with a ZWFS using the light going through the reflective focal plane mask to control low-order aberrations \cite{Pourcelot_2022, Pourcelot_2023}. The latest results show contrast levels down to a few $10^{-8}$ in narrowband and boardband \cite{Soummer_2024}. In this contribution, we briefly recall the main features of the testbed that are relevant for our study with the MOWFS.

The star is simulated with a laser going through an optical fiber at $\lambda=$ 638\,nm. The primary mirror is represented using an outline frame defining the edge of the telescope and a 37-segment PTT111L deformable mirror (DM) from Iris-AO. The later has an inscribed circular diameter of 7\,mm with segments of 1.4\,mm. Each segment is controlled with 3 actuators to generate piston, tip, and tilt (PTT) modes. The maximum amplitude of the actuators is around 5\,$\mu$m, and the system is driven with a 14-bit electronic device. 

\begin{figure}[!ht]
    \centering
    \includegraphics[width=0.6\columnwidth]{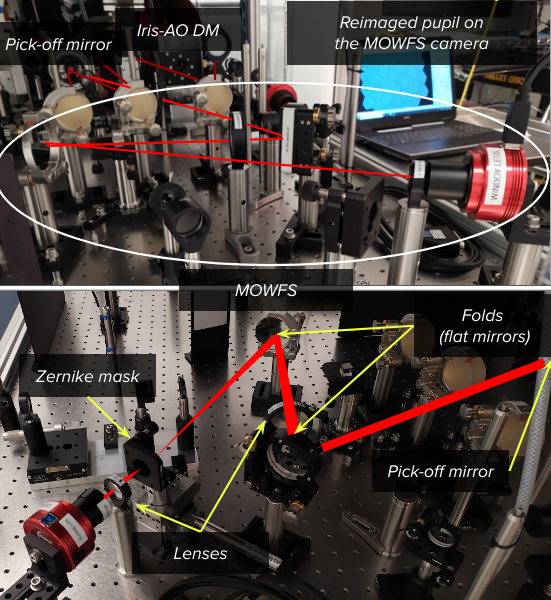}
    \caption{Pictures of the MOWFS setup on the HiCAT testbed during the pre-installation in June 2023.}
    \label{fig:MOWFS}
\end{figure}

On HiCAT, in the current setup, a pick-off mirror has been installed between the Iris-AO DM and the apodizer to send the light either to the science path or to the MOWFS. In the MOWFS configuration, the beam is going through a combination of lenses to form the source image on the ZWFS and the sensor signal on a CCD camera, see Figure \ref{fig:MOWFS}. 

The MOWFS uses a Zernike mask with a diameter of 41\,$\mu$m, which corresponds to 2.5\,$\lambda f / D$ with $f = $ 500\,mm, and $D = $19.2\,mm. The CCD camera is a ZWO ASI178MM with 824x824 pixels. 

A digital twin of HiCAT using the CATKit2 library \cite{Por_2024} is available to numerically reproduce the testbed behaviour and the light propagation along the different elements of the bench. This simulator also enables the development of scripts of experiments before their implementation and exploitation on hardware. In the case of the MOWFS, we have implemented the different parts to drive the pick-off mirror, the Zernike mask alignment and the camera for image acquisition.

\section{MOWFS sensitivity study}

\subsection{Preliminary characterization of the Iris-AO segmented DM} \label{subsec:iris-ao}

Our main objective is to develop a wavefront control loop to stabilize a dark zone in the presence of drift on the segmented primary mirror. To perform such a study, the characterization of the Iris-AO DM on the HiCAT testbed is required. In this section, we present the preliminary tests and results to measure the quantization of the discretization steps of the DM actuators.

\begin{figure}[!ht]
    \centering
    \includegraphics[width=0.495\columnwidth]{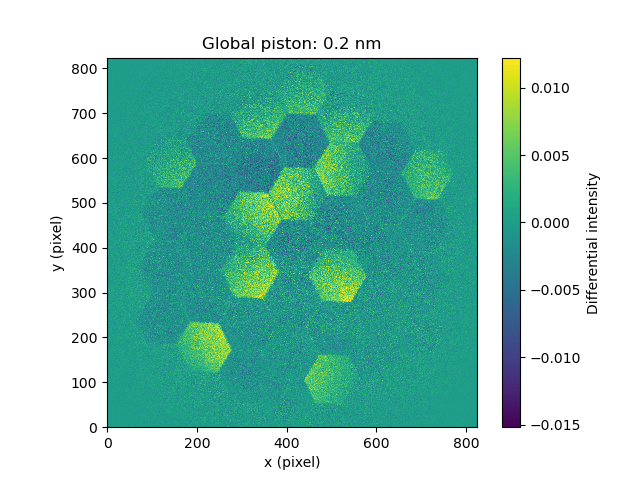}
     \includegraphics[width=0.495\columnwidth]{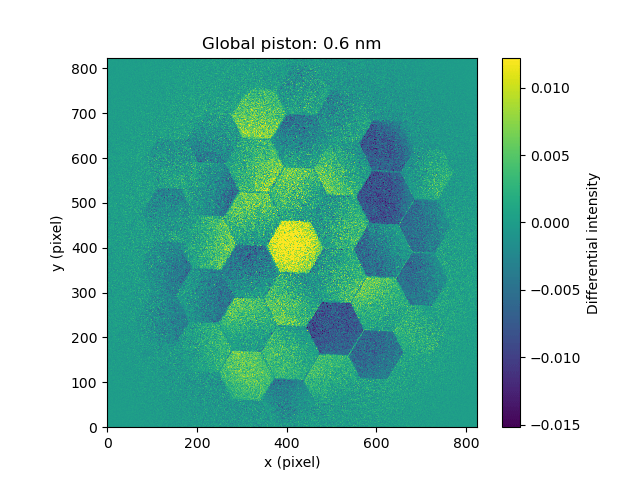}
    \caption{Differential intensity on the MOWFS for a global piston command of 0.2 and 0.6\,nm RMS (left and right). Instead of having a flat intensity map due to the global piston, different values are observed on each segment, showing a non-uniform response of the Iris-AO DM actuators.}
    \label{fig:gif_steps_frames}
\end{figure}

As a first experiment, we start with a flat map introduced on the DM to measure a reference signal on the MOWFS. We then apply a global piston with a given value for all the segments of the DM and we determine the response of the MOWFS by performing a differential intensity measurement. In our procedure, we take measurements by switching from the reference image to the image for a global piston in an alternative way for two reasons. First, we want to increase the signal to noise ratio (SNR) by stacking the differential images over 10 iterations. The second reason is related to a mechanical instability of the mount of the pick-off mirror and some turbulence associated to the motor of this mount. Figure \ref{fig:gif_steps_frames} shows the differential intensity for a global piston of 0.2 and 0.6\,nm RMS.

For each global piston, the intensity map shows a different response for each of the segments. This behaviour might be related to the difference in terms of actuator response for a given command. Qualitatively, the overall shape is not the same for both global piston commands, showing possible different dynamic from one segment to another. It is therefore important to characterize the response of each actuator accurately.

\begin{figure}[!ht]
    \centering
    \includegraphics[width=0.67\columnwidth]{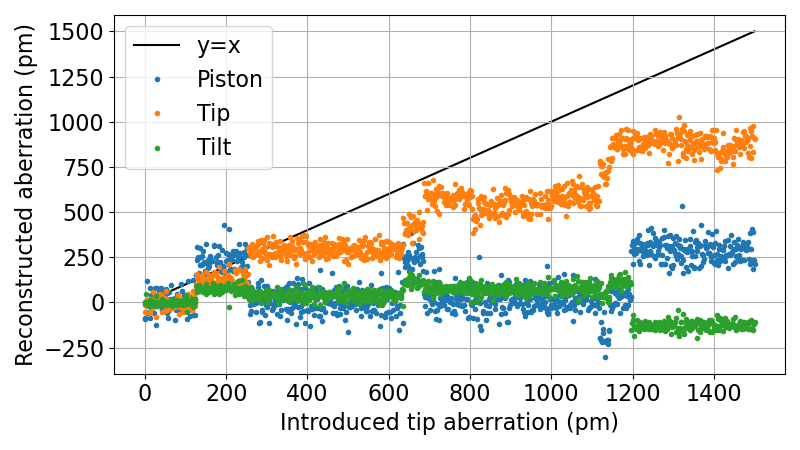}
    \caption{MOWFS response in PTT for a tip mode introduced on the central segment of the Iris-AO DM.}
    \label{fig:DM_steps}
\end{figure}

As a preliminary calibration, we analyze the MOWFS response in PTT for a single mode introduced on the central segment, by using a ramp of tip from 0 to 1500\,pm, with a sampling of 1.5\,pm, see Figure \ref{fig:DM_steps}. The acquisition procedure is the same as previously, i.e differential image between a reference image and an image with the introduced mode. The integration time for each data point is 35\,ms.

At first sight, we notice two things. First, the response curves exhibit the presence of steps, instead of a slope as expected. This shape might be due to the 14-bit voltage resolution of the electronics, leading to the discretization of the DM command. Secondly, for an introduced tip, we measure a combination of PTT modes. This effect is most likely related to the non-uniform response of the actuators for a given mode. These effects can represent a limiting factor for the introduction and correction of PTT in the context of the dark zone stability in the presence of drift on the telescope segments.

For a given tip with an amplitude ranging from 250 to 630\,pm and from 680 to 1129\,pm, the measured piston and tilt are approximately null. In this configuration, the trend of the measured tip does not follow the unitary slope. Work is in progress to understand the origin of this discrepancy. Possible explanations are: calibration errors due to the inaccurate alignment of the ZWFS mask with respect to the source position at focus, the presence of turbulence due to the heating coming from the motor mount of the pick-off mirror, and the non-uniform response of the DM actuators.

\begin{figure}[!ht]
    \centering
    \includegraphics[width=0.495\columnwidth]{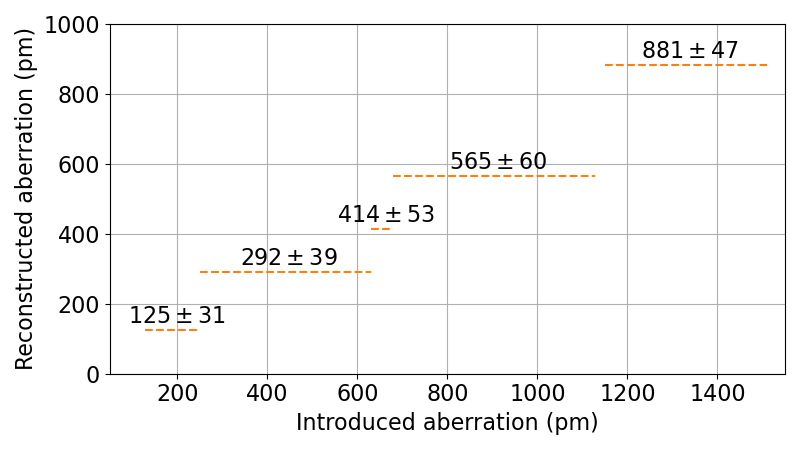}
    \includegraphics[width=0.495\columnwidth]{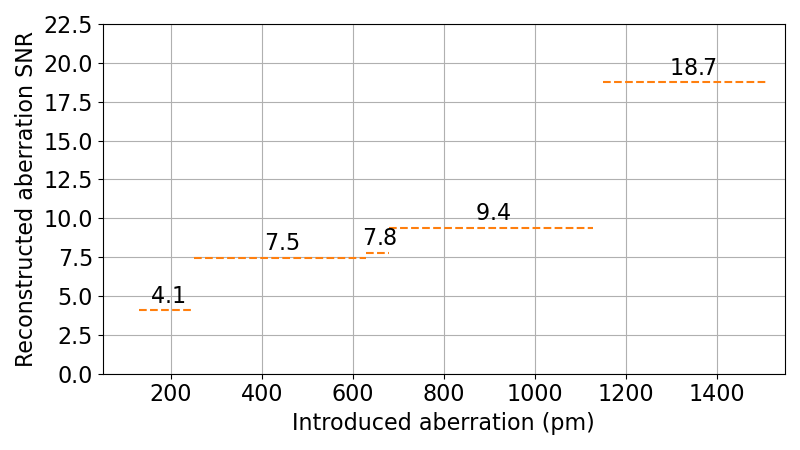}
    \caption{Mean values with RMS (left) and SNR (right) for the discretization steps as a function of the introduced tip.}
    \label{fig:DM_SNR}
\end{figure}

Some information can still be extracted from these steps, for instance, the noise level and the SNR for the characterization of the Iris-AO DM response. As the values of the steps are very stable, we can associate them with the temporal evolution of a segment mode with a specific amplitude. We can then derive the mean value, but also the RMS of the steps, which corresponds to the noise level of the measurements. This information allows us to determine the SNR at which the discretization steps are measured with the MOWFS, see Figure \ref{fig:DM_SNR}.

The left plot shows that the MOWFS allows us to measure a minimal step of 125$\pm$31\,pm. For this discretization step, the right plot shows that we reach a SNR = 4, showing the ability of our MOWFS to characterize the Iris-AO DM discretization steps at a sub-nanometric level with a high SNR. In the same plot, we also notice an increase of the SNR with the introduced amplitude. This result is consistent with our expectations as we are working in a high source flux regime and we are not limited by photon noise or read-out noise.

\subsection{Characterization of the MOWFS capabilities}

The ability to measure wavefront errors down to a few tens of picometers is a key aspect for exo-Earth observations with future large space observatories. To probe wavefront errors at different spatial frequencies, HiCAT uses several wavefront sensors such as the MOWFS. This section aims to characterize the wavefront error detection limit of this sensor on the testbed in air for a source signal with a given SNR. 

\begin{figure}[!ht] 
    \centering
    \includegraphics[width=0.67\columnwidth]{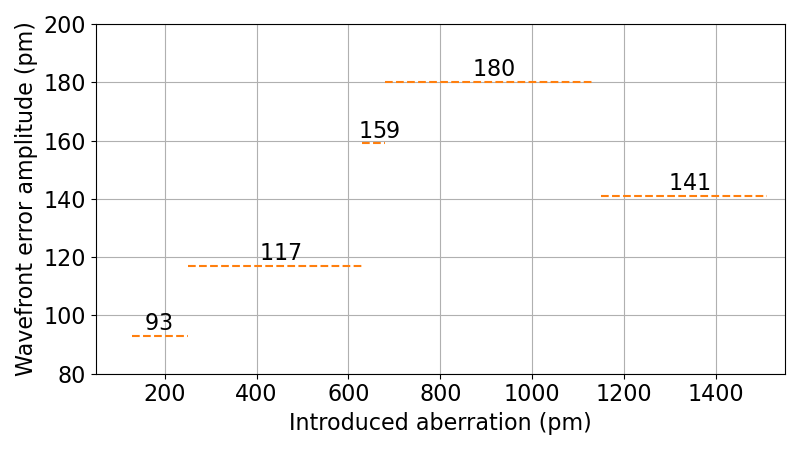}
    \caption{Expected measurement of the wavefront error amplitude for a tip mode on the central segment with the MOWFS and a SNR of 3.}
    \label{fig:DM_SNR=3_mean}
\end{figure}

To reach this goal, we rely on the smallest Iris-AO discretization steps that have been measured by the MOWFS in the previous section. For the steps with an amplitude of 125 and 292\,pm, the MOWFS shows a measurement RMS of 31 and 39\,pm. Figure \ref{fig:DM_SNR=3_mean} shows the wavefront detection limit for a SNR of 3 with the previous discretization steps. For the first and second discretization steps and assuming a SNR of 3, our MOWFS is able to measure a wavefront error with an amplitude down to 93\,pm and 117\,pm RMS. 

Since the second step shows a larger sequence of data points than the first step, we consider these data to determine the temporal evolution of the MOWFS detection limit. We recall that the exposure time for each data point is 35\,ms.

\begin{figure}[!ht]
    \centering
    \includegraphics[width=0.67\columnwidth]{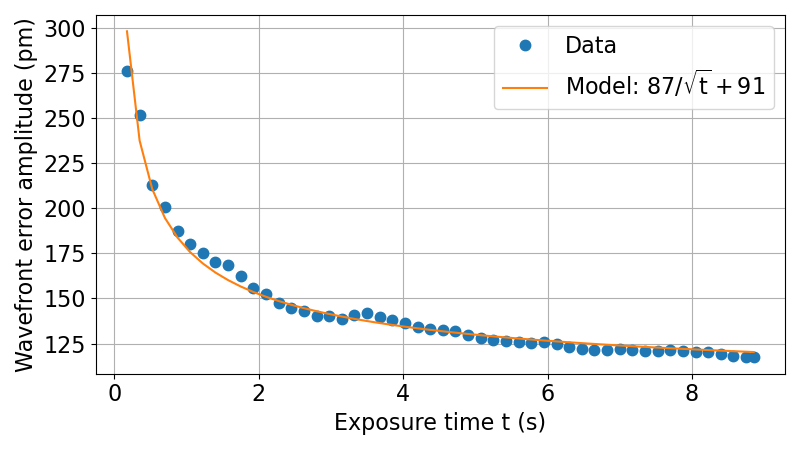}
    \caption{MOWFS detection limit at SNR = 3 for a tip mode on the central segment. The solid line corresponds to a model with a power law fitting the data points.}
    \label{fig:DM_detection}
\end{figure}

We determine the wavefront error detectability of the MOWFS for different integration times. The wavefront error amplitude is computed for an associated number of exposures and scaled with a given SNR to achieve the associated detection limit. In our study we set the SNR to 3. Figure \ref{fig:DM_detection} shows the temporal evolution of the MOWFS detection limit.

The results show a decrease in the mean value of the wavefront error amplitude as we increase the exposure time. We fit the data points with a power law, assuming that we are mostly limited by photon noise. Our resulting model is $87/ \sqrt{t}+91$ in which $t$ denotes the exposure time. Using this model, we can have an averaged estimate of the wavefront error amplitude at longer exposure time and determine the MOWFS detection limit. For an exposure time of 10\,s and 1\,min, the MOWFS reaches a wavefront error amplitude of 119 and 102\,pm. 

Obviously, these results are not direct measurements and present their own limits. However, they provide a good estimate of the MOWFS detection limit on HiCAT in air and in the presence of several calibration errors, as mentioned in section \ref{subsec:iris-ao}. They will be useful to model and extrapolate the metrology capabilities of the wavefront sensors in the context of ultra-stable space telescopes for exo-Earth observations.

\section{Conclusion}

In the context of a segmented aperture space telescope with high-contrast capabilities, a control loop based on a MOWFS is a promising option to ensure the image contrast stability in the presence of fine segment cophasing errors. In June 2023, we installed the MOWFS setup on HiCAT and implemented its numerical version in the digital twin of the bench to run experiments in simulation and on hardware. Our work has led to preliminary studies on the segmented mirror on HiCAT and the MOWFS itself.

In our first study, we have characterized the Iris-AO segmented DM on HiCAT, assessing the amplitude of the DM discretization steps at sub-nanometric levels. Our preliminary study on the central segment showed an estimate of the minimal step for the Iris-AO DM of 125$\pm$31\,pm with a SNR=4. In the forthcoming studies, we will extend our work to all the segments of the Iris-AO DM. Such an analysis will allow us to estimate the minimal level of segment drifts that we could emulate on HiCAT and its impact on the image contrast.  

In a second study, we have determined the detection limit of the MOWFS by analyzing and scaling the measurements of the DM discretization steps. Our results with the data from the second smallest discretization step have showed that our MOWFS should be able to detect aberrations down to 119\,pm and 102\,pm for an exposure time of 10\,s and 1\,min at SNR=3 on the HiCAT testbed in air. These first results on sensor capabilities prove very promising to investigate our ability to control the wavefront errors on HiCAT and therefore, determine the achievable contrast levels with our tools.  

Work is in progress to address some calibration issues related to the Zernike mask misalignment, the turbulence induced by the motor of a pick-off mirror on the beam, and the DM actuator responses. The expected improvements will allow for a better quality, accuracy and reproductibility of the results for the DM characterization and the MOWFS capabilities. One of the possible contrast limitation on HiCAT might be related to the Iris-AO behavior. We plan to study the DM stability by setting the component to a flat position and performing long-time runs with the MOWFS measurements. This will help us to determine the contrast level due to the Iris-AO and investigate our ability to dig a dark hole with a contrast better than $10^{-8}$ on HiCAT. Such a sensitivity analysis will help us to assess the stability and metrology requirements further for HWO with its segmented primary mirror for exo-Earth observations.

\acknowledgments

This work was supported by the Action Spécifique Haute Résolution Angulaire (ASHRA) of CNRS/INSU co-funded by CNES. B.B. acknowledges PhD scholarship funding from Région Provence-Alpes-Côte d’Azur and Thales Alenia Space. B.B. also acknowledges support from Laboratoire Lagrange through the 2024 BQR Lagrange program and from the Lagrange MPO team for the mission to the SPIE conference in Japan.

The HiCAT testbed has been developed over the past 10 years and benefited from the work of an extended collaboration of over 50 people. This work was supported in part by the National Aeronautics and Space Administration under Grant 80NSSC19K0120 issued through the Strategic Astrophysics Technology/Technology Demonstration for Exo-planet Missions Program (SAT-TDEM; PI: R. Soummer), and under Grant 80NSSC22K0372 issued through the Astrophysics Research and Analysis Program (APRA; PI: L. Pueyo). E.H.P. was supported in part by the NASA Hubble Fellowship grant HST-HF2-51467.001-A awarded by the Space Telescope Science Institute, which is operated by the Association of Universities for Research in Astronomy, Incorporated, under NASA contract NAS5-26555. Sarah Steiger acknowledges support by STScI Postdoctoral Fellowship and Iva Laginja acknowledges partial support from a postdoctoral fellowship issued by the Centre National d’Etudes Spatiales (CNES) in France.

% References
\bibliography{report} % bibliography data in report.bib
\bibliographystyle{spiebib_v1} % makes bibtex use spiebib.bst

\end{document}